\documentclass[preprint,11pt]{elsarticle}
\usepackage{algorithmic}

\usepackage{amssymb}

\journal{Physical Communication}

\newcommand{\eeq}{\end{equation}}

\newcommand{\br}{\mbox{\boldmath $r$}}

\newcommand{\bs}{\mbox{\boldmath $s$}}

\newcommand{\bp}{\mbox{\boldmath $p$}}

\newcommand{\bM}{\mbox{\boldmath $M$}}

\newcommand{\bS}{\mbox{\boldmath $S$}}

\newcommand{\bn}{\mbox{\boldmath $n$}}

\newcommand{\bd}{\mbox{\boldmath $d$}}
\newcommand{\bQ}{\mbox{\boldmath $Q$}}

\newcommand{\bI}{\mbox{\boldmath $I$}}

\newcommand{\ds}{\displaystyle}

\newcommand{\K}{{\cal K}}
\newcommand{\G}{{\cal G}}
\newcommand{\N}{{\cal N}_0}
\def\S{{\cal S}}

\newcommand{\beq}{\begin{equation}}

\newtheorem{theorem}{Theorem}

\newtheorem{proposition}{Proposition}
\newtheorem{algorithm}{Algorithm}

\def\IR{{\mathbb R}}
\def\IC{{\mathbb C}}

\begin{document}

\begin{frontmatter}



\title{Potential Games for Energy-Efficient Resource Allocation in Multipoint-to-Multipoint  CDMA Wireless Data Networks}


\author{Stefano Buzzi and Alessio Zappone}

\address{Consorzio Nazionale Interuniversitario per le Telecomunicazioni (CNIT), and \\
University of Cassino,
Via G. Di Biasio, 43; 03043 Cassino (FR),
Italy; \\ e-mail: \{buzzi, alessio.zappone\}@unicas.it.}

\begin{abstract}
The problem of noncooperative resource allocation in a multipoint-to-multipoint cellular network is considered in this paper. The considered scenario is general enough to represent several key instances of modern wireless networks such as a multicellular network, a peer-to-peer network (interference channel), and a wireless network equipped with femtocells. In particular, the problem of
joint transmit waveforms adaptation, linear receiver design, and transmit power control is examined. Several utility functions to be maximized are considered, and, among them, we cite  the received  SINR, and the transmitter energy efficiency, which is measured in bit/Joule, and represents the number of successfully delivered bits for each energy unit used for transmission. Resorting to the theory of potential games, noncooperative games admitting Nash equilibria in multipoint-to-multipoint cellular networks regardless of the channel coefficient realizations are designed. Computer simulations confirm that the considered games are convergent, and show the huge benefits that resource allocation schemes can bring to the performance of wireless data networks.
\end{abstract}


\end{frontmatter}



\section{Introduction}
The scarcity of natural resources, along with climate changes and earth global warming has led the scientific community to take into serious account the issues of energy efficiency and of energy saving. While traditionally these issues have mainly regarded areas such as civil engineering and electricity generation methods, in recent years several disciplines are embracing a green philosophy and energy efficiency concerns are becoming more and more frequent. Communication theory makes no exception, and in the past few years a new hot research topic has been coming out, i.e. the design of communication schemes and resource allocation procedures able to improve the energy efficiency of current communication networks. Indeed, although the energy consumption of wireless communication devices is rather limited as compared, for instance, to that of server farm and/or telephone exchanges, improved energy efficiency results in longer battery life, reduced electromagnetic pollution, and larger amount of data delivered per energy-unit consumed.

Pioneering early works in this area have been \cite{gallager,golay,pierce}; in \cite{gallager} Gallager examined situations in which the primary constraint on the transmitted signal comes from power limitations rather than from bandwidth limitations, and derived the reliability function for binary on-off signaling; the study \cite{golay} is the first wherein it was noticed that in the Gaussian channel the minimum energy per bit is equal to $\N \ln 2$ (with $\N/2$ being the power spectral density of the additive white Gaussian noise), while in \cite{pierce} the capacity per unit cost in bits per joule (or in bits per photon) of the photon counting channel has been derived. A decade later, the study \cite{verdu90}  analyzed the case in which the input alphabet contains a zero-cost symbol, while \cite{meng} has recently analyzed scaling laws for the bit-per-joule capacity of an energy-limited wireless network wherein each node is interested in communication with a randomly selected partner. The aforementioned papers focus on an information-theoretic view of energy efficiency, namely dealing with fundamental energy efficiency limits that can be attained with optimal communication schemes.

A more pragmatic approach is instead adopted in \cite{goodman}, wherein energy efficiency is defined as the ratio
\beq
R \ds \frac{L}{M}\frac{f(\gamma)}{p} \; ,
\label{eq:ee}
\eeq
with $R$ the transmit data rate, $L/M$ the ratio between the payload length and the total length of each data packet, $p$ is the transmit power, and $f(\cdot)$ the efficiency-function, which is an approximation of the probability of error-free reception of a packet of $M$ bits, and which is usually expressed as $f(\gamma)=(1-e^{-\gamma})^M$, with $\gamma$ the received Signal-to-Interference plus Noise Ratio (SINR). The efficiency-function (\ref{eq:ee}) is measured in bit/joule, and represents the number of data bits that are delivered error-free at the receiver for each Joule of energy used for transmission. In \cite{goodman}, a multiuser wireless network with a non-orthogonal access scheme is considered, and game-theoretic tools are used to analyze the case in which each user tunes his transmit power so as to maximize his own energy efficiency. The study \cite{goodman} has been a pioneering work that has paved the way to many interesting subsequent works that have both adopted the energy efficiency definition reported in Eq. (\ref{eq:ee}), and used a game-theoretic approach.
Indeed, game theory, a branch of mathematics studying the interactions among several autonomous subjects with contrasting interests, can be well used in multiuser wireless networks to model the interactions between selfish active users, who are indeed in mutual competition for the available bandwidth and, in general, for the shared available network resources \cite{jsacSI} (see also \cite{SPM} for a collection of survey papers dealing with
the application of game theory in signal processing and communications).
As examples, the
reader is referred to \cite{goodman,nara2,SaraydarPhD}. There, for a multiple
access wireless data network, non-cooperative and cooperative games
are introduced, wherein users choose their transmit powers in order
to maximize their energy efficiency. While the above studies consider the
issue of power control assuming that a conventional matched filter
is available at the receiver, the paper \cite{meshkati}
considers for the first time the problem of joint linear receiver
design and power control so as to maximize the utility of each
user. In particular, it is shown here that the inclusion of receiver design in
the considered game brings remarkable advantages. The approach of \cite{meshkati} is then extended in \cite{jsacbuzzi}, wherein transmitter optimization, i.e. spreading code allocation, is also considered in addition to power allocation and linear receiver choice,  in \cite{buzzimassaro}, wherein a frequency-selective fading channel model is adopted, and in \cite{WLZAP}, wherein the benefic impact of widely-linear processing on energy efficiency is highlighted.  On the other hand, a game-theoretic framework for energy efficiency maximization has been also developed for ultrawideband communication systems in \cite{bacci}.

While several studies and abundance of results are available on non-cooperative resource allocation procedures for the uplink of single-cell data networks (see the aforementioned papers and references therein for a non-exhaustive list), the case in which there are multiple transmitters and receivers has been instead much less investigated,
also due to the fact that several non-cooperative resource allocation games conceived for single-cell systems appear to be no longer convergent to a stable point in systems with multiple receivers.
As notable exceptions, we cite here the work \cite{mandamulti}, wherein, for the uplink of a multi-cell wireless network,  a non-cooperative power control game for energy efficiency maximization is proposed, and the recent paper \cite{menon} (see also \cite[Chap. 5]{mackenziebook}, wherein the same approach is pursued), wherein, resorting to the theory of potential games \cite{potential}, a non-cooperative spreading code allocation algorithm has been proposed, under the assumption that a simple matched filter is used at the receiver. Roughly speaking, in a potential game each change in the utility enjoyed by a given player due to an \emph{unilateral} change of strategy by that player is paired by a similar change in a global function called the potential function. In a potential game, the best response strategy always leads to a Nash equilibrium (NE), and users, by acting selfishly, serve the greater good without knowing it. Potential games are a quite recent discovery for the communications and signal processing scientific community, and very few papers have considered their application to resource allocation problems in this area \cite{menon2,scutari,pot1,pot2}.

It is worth underlining that all the cited papers concentrate on the issue of energy efficiency at the physical and MAC layers of a wireless network, while, on the other hand, a completely fair assessment of the energy efficiency of a cellular system should take into account the whole amount of energy spent to make it work, and in particular the amount of energy needed to implement the considered energy-saving resource allocation schemes. The research in this area is still at a very early stage. Indeed, preliminary results are available in the very recent paper \cite{IshedenEE}, where a parallel fading Gaussian channel is considered. However, applying the approach of \cite{IshedenEE} to the more complex scenario we consider is quite a huge task which is definitely out of the scope of our work.
In this paper thus, in keeping with previous work in this area,
we just focus on the investigation of ``what can be gained'' in terms of energy efficiency by optimizing resource allocation at the physical and MAC layers of a wireless network, deferring the evaluation of the whole energy consumption of a mobile device to future research.

This paper is concerned with a multipoint-to-multipoint wireless data network using a nonorthogonal multiple access strategy such as code division multiple access (CDMA).
The considered model, as will be better explained in the sequel,  is general enough to model the uplink of a  standard multi-cell wireless data network, an ad-hoc network with multiple peer-to-peer links (the so-called interference channel),  and a wireless network equipped with femtocells \cite{Chandra1}.

%

Using \cite{menon} as our departure point, in this paper we make the following contributions:
\begin{itemize}
\item[-]
Using the potential games framework, we propose and analyze several non-cooperative games for joint transmitter and receiver optimization, and aimed at maximizing utility functions strictly related to the signal-to-interference plus noise ratio (SINR).
For such games we analytically prove that the best-response dynamics always converges to a Nash equilibrium (NE).
\item[-]
We propose a non-cooperative joint transceiver optimization and transmit power control game aimed at the maximization of the energy efficiency of each active user. For such a game, we prove that a NE always exists, and provide extensive numerical evidence to support the conjecture that the proposed algorithm always converges to the equilibrium.

\item[-]
We provide extensive numerical results for the aforementioned relevant scenarios, showing that the proposed games outperform competing alternatives, and showing that a judicious adaptation of the transmitted waveform and of the transmit power may lead to a dramatic increase in the energy efficiency of wireless networks.
\end{itemize}

This paper is organized as follows. Next section contains some background material on potential games and a description of the considered multipoint-to-multipoint wireless network. Section III is devoted to the exposition of transmitter and receiver non-cooperative adaptation games for SINR maximization. Section IV considers the problem of energy-efficient non-cooperative power control and transmitter waveform adaptation. Numerical results are discussed in Section V, while brief concluding remarks are given in Section VI.

\section{Preliminaries and system model}
In this section we give brief details on potential games, introduce the general form of the system model for a multipoint-to-multipoint wireless network and formulate the problem statement.

\subsection{Strategic games and potential games}
In its strategic form, a game $\G$  can be described as a triplet $\G=\left[\K, \left\{\S_k\right\}, \left\{u_k\right\} \right]$, wherein $\K$ is the set of players (e.g., the communicating devices in a multiple access network), $\S_k$ is the set of all possible strategies for the $k$-th player, and $u_k$ represents the utility function or payoff of the $k$-th player; $u_k$ is a scalar function depending on the strategies taken by all the players of the game. Thus, a change in strategy from one player affects all other players as well, and triggers a dynamic process, in which players iteratively update their own strategies as a reaction to changes in the strategies of the other players. This process is usually referred to as best-response dynamics, since in each iteration, given the strategies of the other players, each player responds by choosing the strategy that maximizes his own utility function.
The main question here is to understand whether an equilibrium point can be reached or if the players indefinitely go on changing their strategies in a restless fashion. A key concept is thus the notion of
 {\em Nash equilibrium} (NE) . Let
$$
(s_1, s_2, \ldots, s_K) \in \S_1 \times \S_2 \times \ldots \S_K
$$
denote a certain strategy $K$-tuple for the active users. The point $(s_1, s_2, \ldots, s_K)$ is an NE if for every user $k$, we have
$$u_k(s_k, \bs_{-k}) \ge u_k(s_k^*, \bs_{-k})\; ,
$$
$\forall s_k^* \neq s_k\, .$, wherein the vector $\bs_{-k}$, as customary in the game-theoretic literature, denotes the vector of the strategies of all the users but the $k$-th one.
Otherwise stated, at a NE, no user can {\em unilaterally} improve its own utility by taking a different strategy. Thus, at a NE, each user, provided that the other users' strategies do not change, is not interested in changing his own strategy.

We give now the formal definition of a potential game \cite{potential}. A strategic game $\G=\left[\K, \left\{\S_k\right\}, \left\{u_k\right\} \right]$ is called
an \emph{exact potential game} if there exists a function $V : \S_1 \times \S_2 \times \ldots \S_K \rightarrow {\cal R}$ such that for any $k \in \K$
and for any $(s_k, \bs_{-k}), (s_k^*, \bs_{-k}) \in \S_1 \times \S_2 \times \ldots \S_K$, we have
\beq
u_k(s_k, \bs_{-k}) - u_k (s_k^*, \bs_{-k})= V(s_k, \bs_{-k}) - V(s_k^*, \bs_{-k}) \; .
\eeq
Likewise, the game $\G$ is an \emph{ordinal potential game} if the aforementioned function $V$ is such that
\beq
u_k(s_k, \bs_{-k}) > u_k (s_k^*, \bs_{-k}) \Rightarrow  V(s_k, \bs_{-k}) > V(s_k^*, \bs_{-k}) \; .
\eeq
The function $V$ is called the exact (respectively, ordinal) potential of the game.

In an exact potential game, Nash equilibria include maximizers of the potential function (note that generally, the reverse is not true), and, if the utility functions are continuous and the strategy spaces are compact, the best response dynamics will converge to an NE of the game. This interesting feature follows from the fact that each time a player plays his best response, thus maximizing his utility function, he is also increasing the potential function. We remark that convergence holds also in the case in which the players update their strategy one at a time at random epochs (note that a particular case is that in which the updates are performed in a round-robin fashion). It is worthwhile to note that each player can also compute his strategy at the NE by separately implementing (possibly off-line) the entire best-response dynamics.
Roughly said, in a potential game wherein the potential function is bounded from above, any best response dynamic will always converge to an NE: this is a very attractive property that can be used, as we will be showing in the sequel of the paper, to obtain convergent noncooperative games.

\subsection{System model for a general multipoint-to-multipoint network}
We consider a multipoint-to-multipoint wireless communication system with $K$ transmitters and $B$ receivers. The considered multiple access technique is a generic non-orthogonal scheme, such as direct-sequence CDMA; we denote by $N$ the system processing gain. Let $h_{i,j}$ be the real channel gain between the $i$-th transmitter and the $j$-th receiver; moreover, denote by $a(i)$ the index of the receiver assigned to the $i$-th transmitter\footnote{Note that we are assuming here that each user is assigned to a certain receiver, i.e. communicating groups have already been formed.}. After chip-matched filtering and chip-rate sampling, the $N$-dimensional received data vector at the $\ell$-th receiver, say $\br_\ell$, can be written as\footnote{Since we are sampling at the chip-rate, the received waveform, as observed in the signaling interval, is converted into an $N$-dimensional vector.}
\beq
\br_\ell = \ds \sum_{k=1}^K \sqrt{p_k} h_{k, \ell} b_k \bs_k + \bn_\ell \; , \quad \ell=1, \ldots, B \; .
\label{eq:rell}
\eeq
where $b_{k}$, $p_{k}$, and $\bs_{k}$ are the $k$-th transmitter's unit-modulus information symbol, transmitted power, and unit-norm spreading code, respectively, while $\bn_{\ell}$ is the thermal noise at the $\ell$-th receiver, modeled as a zero-mean, white Gaussian process with covariance matrix $\sigma^{2}\bI_{N}$.
Note that the above model is general enough to include the following relevant scenarios.
\begin{itemize}
\item[-] \emph{Uplink of a Multi-cell wireless network}: If we assume that the $B$ receivers are base stations (BSs) and/or access points (APs), and that the transmitters are mobile users, model (\ref{eq:rell}) does represent the uplink of a multicellular network, like, for instance, the one considered in \cite{mandamulti}.
\item[-] \emph{Uplink of a Multi-cell wireless network with femtocells}: Femtocells, also known as home base stations \cite{Saunders2009,Chandra1,Chandra2}, are low-power base stations aimed at indoor coverage of a small number of users. Since they are located very close to the mobile terminals that they serve, their adoption has been shown to dramatically reduce the interference in the network and to increase the overall network throughput due to the possibility of spatial reuse of spreading codes and/or frequencies. It is anticipated that femtocells will play a key role in future wireless networks. Since a system equipped with femtocells is a particular instance of a multi-cell wireless network, model (\ref{eq:rell}) can be used to assess the impact that the use of femtocells may have on the network performance.
\item[-] \emph{Peer-to-peer networks (interference channel)}: If we assume that the number $K$ of transmitters equals the number of receivers $B$, and that each transmitter is coupled with one receiver, we have a set of interfering peer-to-peer wireless links, a scenario that is commonly referred to as interference channel.
\end{itemize}

\subsection{Problem statement}
Given the model (\ref{eq:rell}), we are interested in non-cooperative spreading code adaptation and transmit power control tuning so as to optimize the wireless network energy efficiency. For the sequel, we assume  that a linear detector is used at the receiver, so that the symbol $b_k$ is detected according to the rule
\beq
\widehat{b}_k=\mbox{sgn}\left\{ \bd_k^T \br_{a(k)}\right\} \; ,
\label{eq:decrule}
\eeq
with $\bd_k$ the detection vector for the $k$-th user. Note that the estimate of the symbol $b_k$ is obtained by processing the data $\br_{a(k)}$ only: otherwise stated for the sake of simplicity, we do not take into account soft-handover situations wherein the data transmitted by a given source are decoded at multiple receiver sites; note however that the subsequent derivations may be easily extended to such a scenario.

Given the decision rule (\ref{eq:decrule}),
the SINR  for the $k$-th user is expressed as
\beq
\gamma_k= \ds \frac{p_k h_{k,a(k)}^2 (\bd_k^T \bs_k)^2}{\bd_k^T\left(\sigma^2_n \bI + \ds \sum_{j\neq k}p_j h_{j,a(k)}^2\bs_j \bs_j^T\right)
\bd_k} \; .
\eeq

\section{Non-cooperative transmitter waveform adaptation}\label{Sec:SpreadingCodeDesign}
We start by considering the problem of non-cooperative spreading code adaptation for the maximization of SINR-related utilities, reviewing some already available procedures and introducing new ones, with improved performance. In all of the games in this section, for all $k=1,\ldots,K$, the $k$-th player's strategy set is taken as ${\cal S}_{k}=\{\bs_{k}\in\IC^{N}:\|\bs_{k}\|=1\}$. In other words, the users's spreading codes are constrained to have unit-norm.
Note that adapting the transmit waveform so as to maximize SINR may enable, for a given target quality-of-service, reduction of the transmit power, thus leading to increased energy efficiency. The material of this section is also a prerequisite for the maximization of the energy efficiency utility function (\ref{eq:ee}) of the next Section.

\subsection{Greedy spreading code allocation with LMMSE reception \cite{greedyIA}}\label{Sec:GreedyIA}
We start by reviewing the greedy spreading code allocation procedure of \cite{greedyIA}, and proving a theorem about its convergence in multipoint-to-multipoint networks.
Consider the case in which a linear minimum mean square error (LMMSE) filter is used at the receiver. In this case, through standard linear algebra,  the $k$-th user SINR can be expressed as
\beq
\gamma_k=p_k h_{k,a(k)}^2 \bs_k^T \left( \sigma_n^2 \bI + \sum_{j \neq k}p_j h_{j,a(k)}^2 \bs_j \bs_j^T \right)^{-1}\bs_k \; .
\label{eq:greedyIA}
\eeq
Given the above expression, it is trivially shown that the SINR-maximizing spreading code for the $k$-th user is the eigenvector associated to the minimum eigenvalue of the matrix
$$\left( \sigma_n^2 \bI + \sum_{j \neq k}p_j h_{j,a(k)}^2 \bs_j \bs_j^T \right)\; ,$$
which is indeed the covariance matrix of the overall interference suffered by the $k$-th user.
The non-cooperative game wherein users cyclically update their spreading code in order to maximize the SINR in
Eq. (\ref{eq:greedyIA}) is widely known as greedy interference avoidance procedure \cite{greedyIA}.
Such a procedure is known to be always convergent in single-cell systems, and in the following proposition we prove that it always converges in underloaded multipoint-to-multipont systems.

\begin{proposition}\label{Prop:GreedyIA_Under}
Consider the system model of Eq. (\ref{eq:rell}), and assume $K\leq N$. Then, the greedy interference avoidance algorithm always converges to a set of orthonormal signatures.
\end{proposition}

\noindent
{\bf Proof.} Assume the spreading codes are initialized to $\bs_{1}^{0},\ldots,\bs_{K}^{0}$. If $K \leq N$, the minimum eigenvalue of the matrix
$$\left( \sigma_n^2 \bI + \sum_{j \neq k}p_j h_{j,a(k)}^2 \bs_j \bs_j^T \right)\; ,$$
is $\sigma^{2}$, since the matrix $\sum_{j \neq k}p_j h_{j,a(k)}^2 \bs_j \bs_j^T$ is not full-rank. Note that for any set of spreading codes, each user can compute a unit-norm eigenvector corresponding to $\sigma^{2}$ by choosing a spreading code orthogonal to the spreading codes of the other users. Then, after one signature update for all users, we are left with a set of orthonormal spreading codes, and convergence is reached.
\hfill \rule{2mm}{2mm}

\medskip

Unfortunately, the greedy interference avoidance algorithm does not always converge in overloaded multipoint-to-multipoint systems. However, for comparison purposes in the forthcoming Section V on numerical results we will include performance results for this technique as well.

\subsection{Minimization of the individual MSE \cite{ulukusyates,yener}}\label{Sec:greedyMSE}
As an alternative optimization criterion, we can consider minimization of the individual MSE. The MSE for the $k$-th user, say $\epsilon^2_k$,  is expressed as
\beq
\begin{array}{cc}
\epsilon_k^2=E\left\{ (b_k-\bd_k^T \br_{a(k)})^2\right\} =
1 -2 \sqrt{p_k} h_{k,a(k)}\bd_k^T \bs_k-  \\ -
\ds \frac{\N}{2}\| \bd_k\|^2 + \bd_k^T\left( \ds \sum_{j=1}^K
p_j h_{j,a(k)}^2 \bs_j \bs_j^T \right) \bd_k \;.
\end{array}
\label{eq:msek}
\eeq
Following \cite{ulukusyates,yener}, it is easily seen that the minimizer of $\epsilon_k^2$ can be obtained as the unique stable fixed point of the following iterations:
\beq
\left\{\begin{array}{llll}
\bd_k=\sqrt{p_k} h_{k,a(k)}\bM_{\br_{a(k)}}^{-1} \bs_k \; , \\
\bs_k= \bd_k /\|\bd_k\| \; ,
\end{array} \right.
\label{eq:iterationsMSE}
\eeq
for any $k=1, \ldots, K$. In the above equation $\bM_{\br_{a(k)}}=E\left\{ \br_{a(k)} \br_{a(k)}^T\right\}$ is the covariance matrix of the data vector received at the $a(k)$-th AP.
Now, it is well known that iterations (\ref{eq:iterationsMSE}) are always convergent in a single-cell system, and using a similar proof as in Proposition \ref{Prop:GreedyIA_Under}, it can be shown that it converges in underloaded multipoint-to-multipoint networks, too. However, convergence is not guaranteed in overloaded multi-cell systems. Again, in the following we will be including performance results also for this technique for comparison purposes.

\subsubsection{A side result: waveform adaptation in the downlink of single-cell networks}
It is worth mentioning that a slight modification of the considered system model permits the analysis of the downlink of a single-cell network.
Indeed, assume that
we have only one transmitter, and $B$ distinct mobile receivers. Denoting by $p$ the transmit power, and by $\{g_{b}\}_{b=1}^{B}$ the channel coefficients from the transmitter to the $B$ receivers, the signal received by the $b$-th receiver is expressed as
\begin{equation}
\ds \br_{b}=\sqrt{p}g_{b}\left(\sum_{\ell=1}^{B}b_{\ell}\bs_{\ell}\right)+\bn_{b}\;,
\end{equation}
while its covariance matrix is $\bM_{b}=pg_{b}^{2}\bS\bS^{H}+\sigma_{n}^{2}\bI$. The SINR achieved by the $b$-th receiver can be expressed as
\beq
\gamma_b= \ds \frac{p g_{b}^{2} (\bd_b^T \bs_b)^2}{\bd_b^T\left(\sigma^2_n \bI + \ds p\sum_{j\neq b} g_{b}^2\bs_j \bs_j^T\right)
\bd_b}\;.
\eeq
In this scenario, iterations (\ref{eq:iterationsMSE}) can be written as
\beq
\left\{\begin{array}{llll}
\bd_b=\sqrt{p} g_{b}\bM_{b}^{-1} \bs_b \; , \\
\bs_b= \bd_b /\|\bd_b\| \; .
\end{array} \right.
\label{eq:iterationsMSEDownlink}
\eeq
Generalizing the proof for the uplink scenario that is reported in \cite{ulukusyates}, it is easy to show that iterations (\ref{eq:iterationsMSEDownlink}) always converge, even if each user has a different received data covariance matrix. We do not give further details here for the sake of brevity.

\subsection{Maximization of the opposite of the sum of inverse SINR \cite{menon}}\label{Sec:Menon}
As previously discussed, non-cooperative maximum SINR game with respect to the spreading code and uplink receiver \cite{greedyIA} is not always convergent in overloaded multi-cell networks. In \cite{menon}, instead, based on the theory of potential games, a modification to the utility function to be considered has been introduced, so as to have a guaranteed convergence for any channel realizations. Let us thus assume that a matched filter (MF) is used at the receiver and
consider the negative sum of the inverse SINR, i.e.:
\beq
V=\ds - \sum_{k=1}^K \ds \frac{1}{\gamma_k}\; .
\eeq
Pointing out the dependence on the $k$-th spreading code $\bs_k$, $V$ can be expressed as
\beq
\begin{array}{lll}
V=-&\bs_k^T \left[\ds \frac{\sigma_n^2}{p_k h_{k,a(k)}^2}\bI  
\ds + \sum_{j\neq k} \left(\frac{p_jh_{j,a(k)}^2}{p_kh_{k,a(k)}^2} +
\frac{p_kh_{k,a(j)}^2}{p_jh_{j,a(j)}^2}\right) \bs_j \bs_j^T \right]\bs_k - D \; ,
\end{array}
\eeq
with $D$ an additive term independent of $\bs_k$. It is thus clear that a non-cooperative game wherein the utility function to be maximized for the $k$-th user is
\beq
\begin{array}{lll}
u_k=- & \bs_k^T \left[\ds \frac{\sigma_n^2}{p_k h_{k,a(k)}^2}\bI 
\ds
+ \sum_{j\neq k} \left(\frac{p_jh_{j,a(k)}^2}{p_kh_{k,a(k)}^2} +
\frac{p_kh_{k,a(j)}^2}{p_jh_{j,a(j)}^2}\right) \bs_j \bs_j^T \right]\bs_k\; ,
\end{array}
\eeq
is a potential game with potential function $V$. Consequently, the associated best-response dynamics will always converge to an NE.

\subsection{Greedy interference avoidance revisited}\label{Sec_IArevisited}
We now propose a new non-cooperative game which will be shown to achieve much superior performance levels than the previously discussed solutions.

Since the greedy interference avoidance procedure is not always convergent in multipoint-to-multipoint systems, we resort to the theory of potential games in order to come up with a modified utility function whose non-cooperative maximization leads to an NE.
First of all, it is well-known that the linear detector that maximizes the individual SINR is the LMMSE detector. Hence, consider the case in which  an LMMSE detector is used at the receiver, so that the $k$-th user SINR can be shown to be written as
$$\gamma_k=p_k h^2_{k,a(k)}\bs_k^T\left(\sigma_n^2 \bI + \sum_{j \neq k}p_j h_{j,a(k)}^2 \bs_j \bs_j^T  \right)^{-1}\bs_k \; .$$
Considering the maximization of the opposite of the sum of the inverse SINRs (as done in \cite{menon} for the case of a matched filter receiver) reveals to be a complicated task in this case, and, also, maximization of the sum of the SINRs  turns out to be complicated as well. Based on intuitive reasoning, we consider instead the following quantity
\beq
\ds
\begin{array}{ccc}
Q= - \ds \sum_{m=1}^K
\rho_m= 
- \sum_{m=1}^K
p_m h_{m,a(m)}^2 \bs_m^T \left( \sigma_n^2 \bI + \sum_{j \neq m}p_j h_{j,a(m)}^2 \bs_j \bs_j^T \right)\bs_m \; .
\end{array}
\eeq
Note that the above quantity leads to a mathematically manageable utility, and is directly tied to the SINRs enjoyed by the active users in the network, since it is easy to show that $Q$ is an increasing function of the SINR of each user.
Upon straightforward algebraic manipulation, we find
\beq
\begin{array}{ccc}
\ds Q= -\sum_{m=1}^K
\rho_m= \\
\underbrace{-\bs_k^T \left[
p_k h_{k,a(k)}^2 \sigma_n^2 \bI + \ds \sum_{j \neq k}p_k p_j h_{k,a(k)}^2  h_{j,a(k)}^2 \bs_j \bs_j^T +
\sum_{j \neq k}p_k p_j h_{k,a(j)}^2  h^2_{j,a(j)} \bs_j \bs_j^T\right] \bs_k}_{{\mbox{depends on }}\,  \bs_k} + \\
\ds \underbrace{-\sum_{j=1, j \neq k}^Kp_j h_{j,a(j)}^2\bs_j^T\left(\sigma_n^2 \bI + \ds \sum_{\ell \neq k,j}
p_l h^2_{\ell, a(j)}\bs_\ell \bs_\ell^T\right)\bs_j}_{{\mbox{does not depend on }}\,  \bs_k}  \; .
\end{array}
\eeq
Accordingly, a non-cooperative game wherein each user aims at maximizing the utility
\beq\label{Eq:UtlilitySINRPot}
\begin{array}{lll}
u_k= & \ds -\bs_k^T \left[\sigma_n^2 \bI + \ds \sum_{j \neq k} p_j   h_{j,a(k)}^2 \bs_j \bs_j^T 
+ \sum_{j \neq k} p_j \ds \frac{h_{k,a(j)}^2}{h_{k,a(k)}^2}  h^2_{j,a(j)} \bs_j \bs_j^T\right]\bs_k \; ,
\end{array}
\eeq
is a potential game whose potential function is $Q$, and therefore it always admits an NE. Moreover, the best-response dynamics in which each player iteratively maximizes (\ref{Eq:UtlilitySINRPot}) converges to an NE of the proposed potential game. For all $k=1,\ldots,K$, maximization of (\ref{Eq:UtlilitySINRPot}) with the constraint $\|\bs_{k}\|=1$ yields the unit-norm eigenvector of
\begin{equation}
\left[\sigma_n^2 \bI + \ds \sum_{j \neq k} p_j   h_{j,a(k)}^2 \bs_j \bs_j^T +
\sum_{j \neq k} p_j \ds \frac{h_{k,a(j)}^2}{h_{k,a(k)}^2}  h^2_{j,a(j)} \bs_j \bs_j^T\right]
\end{equation}
corresponding to the minimum eigenvalue. Moreover, note that following the same arguments as in Proposition \ref{Prop:GreedyIA_Under}, it can be proved that the proposed potential game converges to a set of orthonormal signatures for $K\leq N$, and is therefore equivalent to the greedy interference avoidance procedure in underloaded systems.

\section{A non-cooperative game for energy-efficient communications}
In the previous Section, we have analyzed the problem of non-cooperative maximization of utility functions strictly related to the SINR, which is indirectly tied to the system's energy efficiency\footnote{Indeed, recall that for a given target error probability and/or throughput, waveform adaptation aimed at SINR maximization permits reducing the transmit power, and thus, increases the system energy efficiency.}. In what follows, we resume the energy efficiency of Eq. (\ref{eq:ee}), and consider the joint problem of transmit power control, receiver design and transmit spreading code adaptation for non-cooperative energy efficiency maximization.

As customary in the scientific literature from \cite{goodman} on, the following utility function should be thus considered for the $k$-th user
\begin{equation}\label{Eq:EnergyEfficiency}
u_k=R \ds \frac{L}{M}\frac{f(\gamma_k)}{p_k} \; ,
\end{equation}
with $R$ the transmit data rate, $L/M$ the ratio between the payload length and the total length of each data packet, and $f(\cdot)$ the efficiency-function, which is usually expressed as $f(\gamma_k)=(1-e^{-\gamma_k})^M$. We are here interested in the non-cooperative maximization of $u_k$ with respect to $p_k$, $\bs_k$ and $\bd_k$. Note that
\beq
\ds \max_{p_k, \bs_k, \bd_k} \ds \frac{f(\gamma_k)}{p_k}= \max_{p_k} \ds \frac{f\left(\ds\max_{\bs_k \bd_k} \gamma_k \right)}{p_k}\; .
\label{eq:ee2}
\eeq
In the above equation we have exploited the  fact that $f(\cdot)$ is an increasing function, and that only the numerator (and not the denominator) depends on the vectors $\bd_k$ and $\bs_k$.
Otherwise stated, each user can take care first of SINR maximization with respect to $\bs_k$ and $\bd_k$, and then of the maximization of its energy efficiency, i.e. of the ratio $f(\gamma_k)/p_k$.  In \cite{mandamulti}, it is shown that in a multi-cell network, the non-cooperative maximization of the energy efficiency with respect to the transmit power only, always converges to an NE when a plain matched filter is used at the receiver. Following the same approach as in \cite{mandamulti}, it is easy to show that non-cooperative energy efficiency maximization with respect to the transmit power converges to an NE also in the case in which an LMMSE multiuser detector\footnote{Note that the LMMSE detector is the optimum energy-efficient receive filter, since LMMSE filtering is well-known to maximize the received SINR.} is used at the receiver (details are not given for the sake of brevity, but we will be providing simulation results for this strategy as well). Instead, if spreading code allocation comes into play, things are more involved. Given (\ref{eq:ee2}), individual energy-efficiency maximization should be carried out according to the following algorithm
\begin{algorithm}\label{Alg:Single-CellEE}
\begin{algorithmic}
\STATE
\REPEAT
\STATE STEP 1: for fixed transmit powers, allocate the spreading codes and detection vectors for individual SINR maximization;
\STATE STEP 2: for fixed spreading codes and detection vectors, play the power control game in \cite{mandamulti} for energy efficiency maximization (\ref{Eq:EnergyEfficiency});
\UNTIL{Convergence is reached.}
\end{algorithmic}
\end{algorithm}
While such a procedure always converges for single-cell networks, and indeed results for this scenario are presented in \cite{jsacbuzzi}, convergence is not guaranteed in multipoint-to-multipoint networks because, as it has been pointed out in Section \ref{Sec:GreedyIA}, individual SINR maximization is not always convergent. Thus, in order to obtain an energy efficiency game converging to an NE, spreading codes allocation should not be carried out for SINR maximization. For this reason, we consider a different approach to spreading codes and decoding vectors allocation, as outlined in the following section.

\subsection{Non-cooperative minimization of the TMSE}\label{Sec:PotentialGameTMSE}
In this section, again we use the potential games framework to obtain a convergent non-cooperative game for spreading code and detection vector design.
As a potential function we consider the total MSE, defined as $\sum_{k=1}^K \epsilon^2_k$. Upon some straightforward algebraic manipulations, we have
\beq
\begin{array}{ccc}
\ds \sum_{m=1}^K \epsilon^2_m= \\
\underbrace{1-2\sqrt{p_k} h_{k,a(k)} \bd_k^T \bs_k + \ds \frac{\N}{2}\|\bd_k\|^2+
\bd_k^T\left(\ds \sum_{j=1}^K
p_j h_{j,a(k)}^2 \bs_j \bs_j^T \right) \bd_k +
\ds \sum_{\ell \neq k}\bd_{\ell}^T \left( p_k h_{k,a(l)}^2 \bs_k \bs_k^T\right) \bd_{\ell}}_{{\mbox{depends on }}\,  \bs_k} +
\\
\underbrace{(K-1) + \ds \sum_{\ell \neq k} \bd_{\ell}^T \left( \ds \sum_{j \neq k}p_j h_{j,a(\ell)}^2\bs_j \bs_j^T \right) \bd_{\ell}
-2 \sum_{\ell \neq k}  \sqrt{p_{\ell}} h_{\ell, a(\ell)}\bd_{\ell}^T \bs_{\ell} + \sum_{\ell \neq k}
\ds \frac{\N}{2}\bd_{\ell}}_{{\mbox{does not depend on }}\,  \bs_k} \; .
\end{array}
\eeq
It is easy to realize that the part dependent on $\bs_k$, say $L(\bs_k)$, may be written as
\beq
L(\bs_k)= \epsilon_k^2 + \ds \sum_{\ell \neq k}\bd_{\ell}^T \left( p_k h_{k,a(\ell)}^2 \bs_k \bs_k^T\right) \bd_{\ell}\; ,
\label{eq:L}
\eeq
thus implying that a non-cooperative game in which each player maximizes (\ref{eq:L}) is a potential game whose potential function is the system's TMSE. Consequently, the proposed game always admits an NE, and its associated best-response dynamics will always converge to an NE.
Summing up, we thus consider the following game:
\beq\label{Eq:TMSEgame}
\ds \max_{\bs_k, \bd_k} -L(\bs_k,\bd_{k}) \; \;, \mbox{ subject to:  } \|\bs_k\|^{2}=1\; .
\eeq
Now, maximization of $-L(\bs_k,\bd_{k})$ with respect to $\bd_{k}$ yields the LMMSE receiver,
\beq\label{Eq:BRreceiver}
\bd_{k}=\sqrt{p_{k}}h_{k,a(k)}\left(\sum_{\ell=1}^{K}p_{\ell}h_{\ell,a(k)}^{2}\bs_{\ell}\bs_{\ell}^{H}+\sigma^{2}\bI_{N}\right)^{-1}\bs_{k}\;,
\eeq
whereas applying standard Lagrangian techniques, maximization with respect to $\bs_{k}$ yields
\beq\label{Eq:BRspreading}
\bs_k=\sqrt{p_k} h_{k,a(k)}\left( \lambda_{k} \bI_{N} + \ds \sum_{\ell=1}^K p_k h^2_{k,a(\ell)}\bd_{\ell} \bd^T_{\ell} \right)^{-1} \bd_k \; ,
\eeq
where the Lagrange multiplier $\lambda_{k}$ is to be chosen such that $\|\bs_k\|^{2}=1$. The computational complexity of the above equations is basically ${\cal O}(N^3)$, due to the need for matrix inversion.
Note that in (\ref{Eq:BRspreading}) we have implicitly assumed that the matrix $\lambda_{k} \bI_{N} + \sum_{\ell=1}^K p_k h^2_{k,a(\ell)}\bd_{\ell} \bd^T_{\ell}$ is invertible. Indeed, this is always the case since $\lambda_{k} \bI_{N} + \sum_{\ell=1}^K p_k h^2_{k,a(\ell)}\bd_{\ell} \bd^T_{\ell}$ is singular only if the multiplier $\lambda_{k}$ is equal to the opposite of one of the eigenvalues of $\sum_{\ell=1}^K p_k h^2_{k,a(\ell)}\bd_{\ell} \bd^T_{\ell}$. However, for such values of $\lambda_{k}$, $\|\bs_{k}\|\;\rightarrow\;+\infty$, thus implying that these values of $\lambda_{k}$ will never satisfy the norm constraint. Otherwise stated, for any set of transmit powers, the resulting multiplier $\lambda_{k}$ is always such that the matrix $\lambda_{k} \bI_{N} + \sum_{\ell=1}^K p_k h^2_{k,a(\ell)}\bd_{\ell} \bd^T_{\ell}$ is invertible.
\\
Defining the matrix $\bM=\sum_{\ell=1}^{K}p_{\ell}h_{\ell,a(k)}^{2}\bs_{\ell}\bs_{\ell}^{H}+\sigma^{2}\bI_{N}$, plugging (\ref{Eq:BRreceiver}) into (\ref{Eq:BRspreading}), and elaborating, for all $k=1,\ldots,K$ we obtain the signature update rule as
\beq\label{Eq:SpreadingUpdate}
\bs_{k}(n+1)=p_{k}h_{k,a(k)}^{2}\left(\lambda_{k}\bM(n)+\ds\sum_{\ell=1}^{K}p_{k}h^{2}_{k,a(\ell)}p_{\ell}h^{2}_{\ell,a(\ell)}\bs_{\ell}(n)\bs_{\ell}^{T}(n)\bM^{-T}(n) \right)^{-1}\bs_{k}(n)\;.
\eeq

\bigskip
Now, we can modify Algorithm \ref{Alg:Single-CellEE} replacing STEP 1 with the novel procedure devised in Section \ref{Sec:PotentialGameTMSE}, and formally state our proposed algorithm for energy-efficiency maximization, as follows.
\begin{algorithm}\label{Alg:EnergyEfficiency}
\begin{algorithmic}
\STATE
\REPEAT
\STATE STEP 1: for fixed transmit powers, allocate the spreading codes and detection vectors according to the non-cooperative game for TMSE minimization, as explained in Section \ref{Sec:PotentialGameTMSE};
\STATE STEP 2: for fixed spreading codes and detection vectors, play the power control game in \cite{mandamulti} for energy efficiency maximization (\ref{Eq:EnergyEfficiency});
\UNTIL{Convergence is reached.}
\end{algorithmic}
\end{algorithm}
Unfortunately, we could not theoretically prove that Algorithm \ref{Alg:EnergyEfficiency} always converges for any channel realizations. However, in the following we prove that Algorithm \ref{Alg:EnergyEfficiency} always admits at least one fixed point, and that in the high-SINR regime, the fixed points of Algorithm \ref{Alg:EnergyEfficiency} tend to those of Algorithm \ref{Alg:Single-CellEE}. Moreover, we point out that in all of our numerical simulations Algorithm \ref{Alg:EnergyEfficiency} has \emph{always} converged thus suggesting the conjecture that Algorithm \ref{Alg:EnergyEfficiency} is indeed always convergent.

\subsection{Existence of fixed points}
We will need the following result.
\begin{theorem}[Brouwer's Fixed-Point Theorem]\label{Th:Brouwer}
Let $Z\subseteq\IR^{K}$ be compact and convex, and $F:\;Z\;\longrightarrow\;Z$ a continuos function. Then $F$ admits a fixed point $z\in Z$.
\end{theorem}
Then, the following proposition holds.
\begin{proposition}
Given any realization of channel coefficients $\{h_{k,a(j)}\}_{k,j=1}^{K}$, Algorithm \ref{Alg:EnergyEfficiency} always admits at least one fixed point.
\end{proposition}

\noindent
{\bf Proof.}
Let us denote by $\IC_{1}^{N\times K}$ the set of matrices in $\IC^{N\times K}$ with unit-norm columns. Denote by $\bS_{0}$ the spreading code matrix used as starting point of the sequence control game of STEP 1 of Algorithm \ref{Alg:EnergyEfficiency}, and consider the function
\beq
\tilde{\bS}:\;\bp=p_{1},\ldots,p_{K}\in [0;P_{max}]^{K},\bS_{0}\in\IC_{1}^{N\times K}\longrightarrow \tilde{\bS}(\bp,\bS_{0})=[\tilde{\bs}_{1},\ldots,\tilde{\bs}_{K}]\in \IC_{1}^{N\times K}\;,
\eeq
where $\tilde{\bS}(\bp)$ is the NE\footnote{As shown in Section \ref{Sec:PotentialGameTMSE}, such an equilibrium always exists.} of the sequence control game reached in STEP 1 of Algorithm \ref{Alg:EnergyEfficiency}. Note that even if the NE of such a game is not unique, we are defining $\tilde{\bS}$ as a function and not a multifunction. Indeed, we stress that the output of $\tilde{\bS}$ is not the set of all the NE of the sequence control game, but only the one NE that is obtained by running STEP 1 of Algorithm \ref{Alg:EnergyEfficiency} with the given channel realizations $\{h_{k,a(j)}\}_{k,j=1}^{K}$ and starting point $\bS_{0}$. We also remark that in the first iteration of Algorithm \ref{Alg:EnergyEfficiency}, the starting point $\bS_{0}$ can be chosen at will, whereas in iteration $n$, it is given by the spreading matrix output by iteration $n-1$.
Moreover, we consider the function
\beq
\tilde{\bp}:\;\bS\in \IC_{1}^{N\times K},\;\bp\in[0;P_{max}]^{K}\;\longrightarrow \; \tilde{\bp}(\bS,\bp)=[\tilde{p}_{1},\ldots,\tilde{p}_{K}]\in[0;P_{max}]^{K}\;,
\eeq
where, for all $k=1,\ldots,K$, $\tilde{p}_{k}$ is the best-response of user $k$ to the strategies of the other players in the power control game of STEP 2 of Algorithm \ref{Alg:EnergyEfficiency}. From \cite{mandamulti} we know that, for all $k=1,\ldots,K$
\beq\label{Eq:BRpower}
\tilde{p}_{k}={\rm min}(P_{max},\bar{p}_{k})\;,
\eeq
where $\bar{p}_{k}$ is the power necessary for user $k$ to achieve a SINR $\bar{\gamma}$, with $\bar{\gamma}$ being the unique, positive solution to the equation $xf'(x)=f(x)$, i.e.
\beq\label{Eq:OptPower}
\bar{p}_{k}=\frac{\bar{\gamma}}{h_{k,a(k)}^{2}\bs_{k}^{T}\left(\sum_{\ell\neq k}p_{\ell}h_{\ell,a(k)}^{2}\bs_{\ell}\bs_{\ell}^{H}+\sigma^{2}\bI_{N}\right)^{-1}\bs_{k}}\;.
\eeq
Algorithm \ref{Alg:EnergyEfficiency} admits a fixed point if the composition
\beq
F:\;\bp\in [0;P_{max}]^{K}\;\longrightarrow\;F(\bp)=\tilde{\bp}(\tilde{\bS}(\bp,\bS_{0}),\bp)\in [0;P_{max}]^{K},
\eeq
admits a fixed point, i.e. if there exists a vector of powers $\bp^{*}$ such that $F(\bp^{*})=\tilde{\bp}(\tilde{\bS}(\bp^{*},\bS_{0}),\bp^{*})=\bp^{*}$.
Since $[0;P_{max}]^{K}$ is a convex and compact subset of $\IR^{K}$, if we can prove that $F$ is a continuous mapping, we can apply Theorem \ref{Th:Brouwer} to obtain the thesis. To this end, from (\ref{Eq:OptPower}) it is seen that, for all $k=1,\ldots,K$, (\ref{Eq:BRpower}) is continuous\footnote{Indeed, recall that when $\bS\in\IC_{1}^{N\times K}$, the spreading codes are constrained to have unit-norm, and hence the denominator of (\ref{Eq:OptPower}) can never be zero.} for any $\bp$ and $\bS\in\IC_{1}^{N\times K}$, thus implying that $\tilde{\bp}$ is continuous in $\bp$, for any given spreading matrix.
Then, $F$ is continuous if the internal function $\tilde{\bS}(\bp,\bS_{0})$ is continuous. To see this, consider the spreading code update function $f$ of the sequence control game, i.e. the function that takes as input a unit-norm spreading code $\bs_{k}$ and outputs another unit-norm spreading code according to equation (\ref{Eq:SpreadingUpdate}). Clearly, $f$ is a continuous function, since it is obtained as the composition of (\ref{Eq:BRreceiver}) and (\ref{Eq:BRspreading}), which both define continuous functions.
Now, the NE of the sequence control game is always obtained by iteratively applying the function $f$ a finite number of times, say $\tilde{n}$. Thus, the function $\tilde{\bS}$ is equivalent to composing $f$ with itself $\tilde{n}$ times, and since the composition of a finite number of continuous functions results in a continuous function, we obtain the thesis.
\hfill
\rule{2mm}{2mm}

\subsection{High SINR regime}
In this section we turn to the analysis of Algorithm \ref{Alg:EnergyEfficiency} in the high SINR regime. We will show that for high SINRs STEP 1 of Algorithm \ref{Alg:EnergyEfficiency} reduces to a game for individual SINR maximization, thus implying that the fixed points of Algorithm \ref{Alg:EnergyEfficiency}, are the NE for the original energy efficiency game in which each player selfishly maximizes (\ref{Eq:EnergyEfficiency}).
To begin with, note that the MMSE filter for the $\ell$-th user can be written as
\begin{equation}
\begin{array}{lllll}
\ds \bd_{\ell}&=\sqrt{p_{\ell}}h_{\ell,a(\ell)}\left(\sum_{j=1}^{K}p_{j}h_{j,a(\ell)}^{2}\bs_{j}\bs_{j}^{T}+\sigma_{n}^{2}\bI_{N}\right)^{-1}\bs_{\ell}
 \\ &
=
\sqrt{p_{\ell}}h_{\ell,a(\ell)}\left(p_{k}h_{k,a(\ell)}^{2}\bs_{k}\bs_{k}^{T}+\bQ_{k,\ell}\right)^{-1}\bs_{\ell}\;,
\end{array}
\end{equation}
where $\bQ_{k,\ell}=\sum_{j\neq k}^{K}p_{j}h_{j,a(\ell)}^{2}\bs_{j}\bs_{j}^{T}+\sigma_{n}^{2}\bI_{N}$ is the interference-plus-noise covariance matrix of the $k$-th user, at receiver $a(\ell)$. Now, applying the matrix inversion lemma, we obtain
\begin{equation}
\bd_{\ell}=\sqrt{p_{\ell}}h_{\ell,a(\ell)}\frac{\bQ_{k,\ell}^{-1}\bs_{\ell}}{1+p_{k}h_{k,a(\ell)}^{2}\bs_{k}^{T}\bQ_{k,\ell}^{-1}\bs_{k}}\;,
\end{equation}
and hence we can write
\begin{equation}\label{eq:RecCorr}
\bd_{\ell}^{T}\bs_{k}=\sqrt{p_{\ell}}h_{\ell,a(\ell)}\left[\frac{\bs_{\ell}^{T}\bQ_{k,\ell}^{-1}\bs_{k}}{1+p_{k}h_{k,a(\ell)}^{2}\bs_{k}^{T}\bQ_{k,\ell}^{-1}\bs_{k}}\right]
\end{equation}
Then, plugging (\ref{eq:RecCorr}) into the second summand of the TMSE in (\ref{eq:L}), we obtain
\begin{equation}\label{Eq:LargeSINRUtility}
\begin{array}{ccc}
\ds  p_{k}\sum_{\ell\neq k}h_{k,a(\ell)}^{2}\left(\bd_{\ell}^{T}\bs_{k}\right)^{2} = p_{k}\sum_{\ell\neq k} h_{k,a(\ell)}^{2}
p_{\ell}h_{\ell,a(\ell)}^{2}\left[\frac{\bs_{\ell}^{T}\bQ_{k,\ell}^{-1}\bs_{k}}
{1+p_{k}h_{k,a(\ell)}^{2}\bs_{k}^{T}\bQ_{k,\ell}^{-1}\bs_{k}}\right]^{2}=\\
 \ds p_{k}\sum_{\ell\neq k} h_{k,a(\ell)}^{2}p_{\ell}h_{\ell,a(\ell)}^{2}\left[\frac{\bs_{\ell}^{T}\bQ_{k,\ell}^{-1}\bs_{k}}{1+\gamma_{k,a(\ell)}}\right]^{2}
\end{array}
\end{equation}
where $\gamma_{k,a(\ell)}$ is the SINR of the $k$-th user at the $\ell$-th access point. As a consequence, for large SINR, we see that (\ref{Eq:LargeSINRUtility}) vanishes, thus implying that the utility function (\ref{eq:L}) can be approximated by $\epsilon_{k}^{2}$, the MSE of the $k$-th user. As a consequence, in this scenario, the game for non-cooperative TMSE minimization that is implemented in STEP 1 of Algorithm \ref{Alg:EnergyEfficiency}, is also a convergent game for individual MSE minimization. Then, recalling that in CDMA systems the $k$-th user's MSE and SINR are linked by the following formula \cite{PoorWangBook}
\beq
\epsilon_{k}^{2}=\frac{1}{1+\gamma_{k}}\;,
\eeq
it is easy to see that individual MSE minimization is equivalent to individual SINR maximization, thus implying that STEP 1 of Algorithm \ref{Alg:EnergyEfficiency} is a convergent game for individual SINR maximization, too. Hence, Algorithm \ref{Alg:EnergyEfficiency} is equivalent to Algorithm \ref{Alg:Single-CellEE}.
Moreover, in this scenario it can be shown that the energy-efficiency game is a separable game \cite{sung}, and similarly to \cite{jsacbuzzi,WLZAP} it can be proved that it admits a unique NE, which is also Pareto-efficient in underloaded systems.


The remarkable performance advantage that the proposed strategy brings with respect to competing resource allocation methods is addressed in the following section.

\section{Numerical results}
In what follows we give extensive simulation results showing the merits of the proposed non-cooperative resource allocation procedures. As anticipated, we have considered three interesting instances of a multipoint-to-multipoint wireless network, as detailed in the following.
We set the processing gain to $N=8$. Users' location have been randomly generated in a square of $10^6$ sq. meters, while the channel coefficients $h^2_{i,j}$ have been generated according to an exponential distribution with mean equal to $d_{i,j}^{-2}$, with $d_{i,j}$ the distance between the $i$-th user and the $j$-th access point. The results that we present have been obtained through averaging over $1000$ independent realizations of the channel coefficients, users' locations, and starting set of spreading codes.

\subsection{Peer to Peer channel (Interference channel)}
\begin{figure}[t]
\centering
\includegraphics[width=13cm]{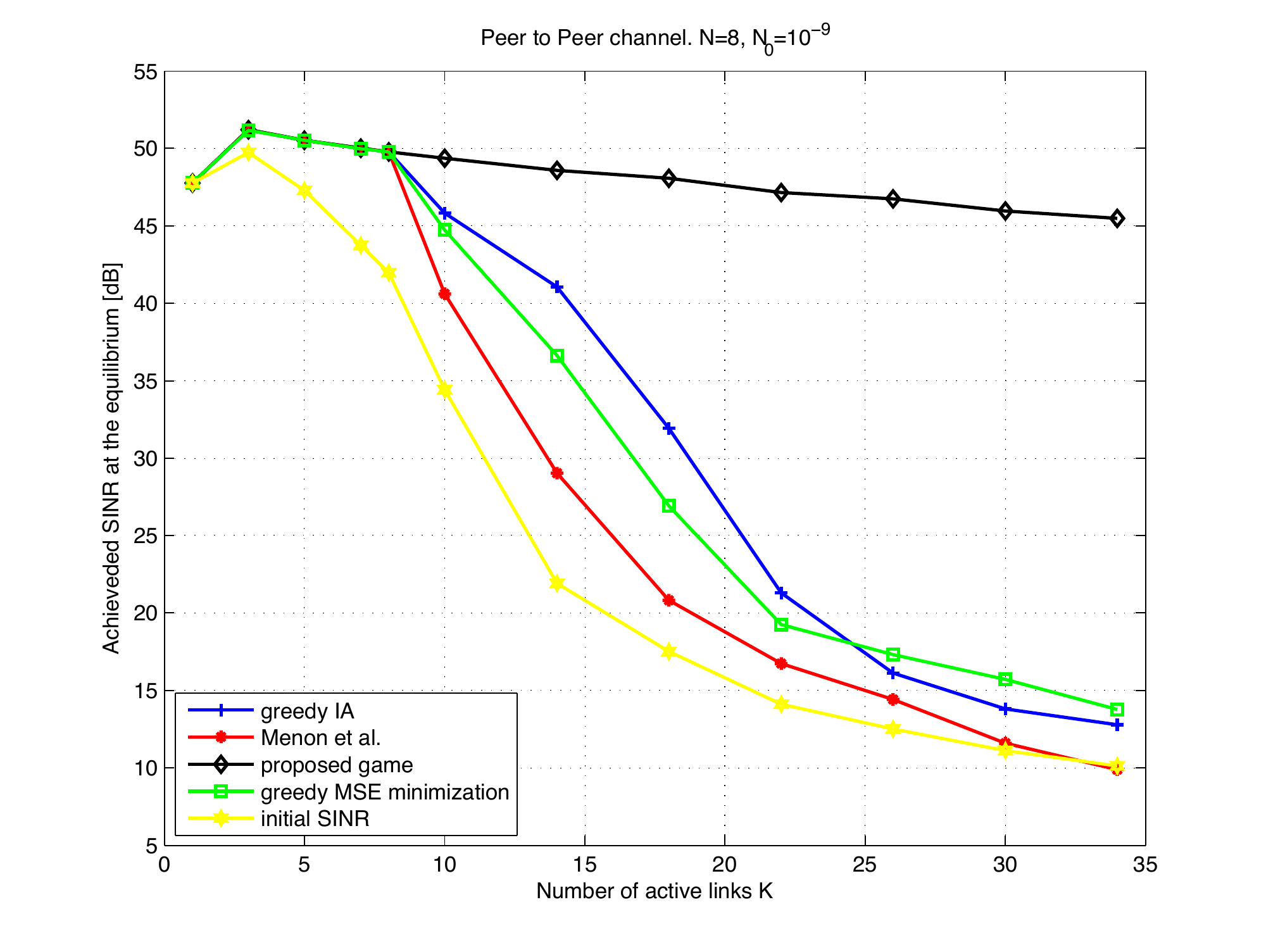}
\caption{Peer to Peer channel. $N=8$. Achieved SINR at the NE for the considered spreading code allocation procedures versus the number of active users.}\label{Fig1}
\end{figure}

\begin{figure}[t]
\centering
\includegraphics[width=13cm]{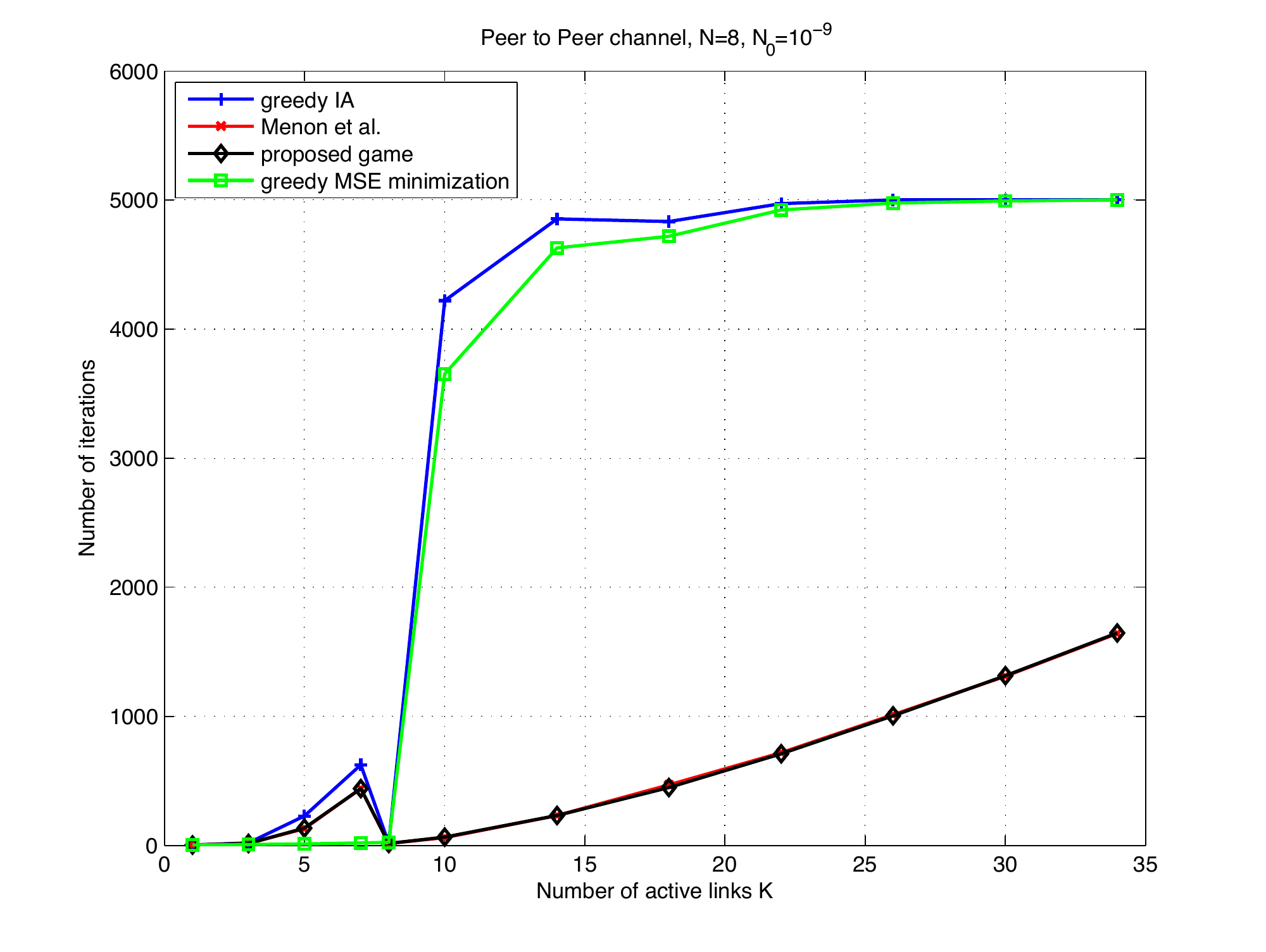}
\caption{Peer to Peer channel. $N=8$. Number of iterations needed to reach the NE for the considered spreading code allocation procedures versus the number of active users. A maximum of 5000 iterations has been included in the simulation program.}
\label{Fig2}
\end{figure}

We considered a DS/CDMA peer to peer channel with $K=B$ active links. It is assumed that the distance between each transmitter and the corresponding receiver is not larger than half of the cell side, i.e. 500 meters. In Figs. \ref{Fig1} and \ref{Fig2}, spreading code allocation is addressed by confronting the following algorithms.
\begin{enumerate}
\item
The proposed spreading code allocation algorithm (Sec. \ref{Sec_IArevisited}).
\item
The algorithm proposed in \cite{menon} (Sec. \ref{Sec:Menon}).
\item
The greedy interference avoidance (Sec. \ref{Sec:GreedyIA}).
\item
The greedy MSE minimization (Sec. \ref{Sec:greedyMSE}).
\end{enumerate}
Since the last two procedures are not always convergent in the considered scenario, in our simulations the maximum number of iterations has been limited to $5000$. Fig. \ref{Fig1} shows the achieved SINR at the NE versus the number of active links.
We also report the initial SINR that is obtained when random spreading codes and MMSE detection vectors are employed. It is seen that for $K\leq N$ all the algorithms have similar performance, thus confirming that in underloaded systems all algorithms converge to a set of orthonormal signatures, and are therefore equivalent. Instead, in overloaded scenarios the proposed procedure largely outperforms all competing alternatives, thus allowing a huge energy conservation for a given target quality-of-service.
In Fig. \ref{Fig2} the number of iterations needed for convergence is plotted versus the number of active links. It is seen that the proposed procedure and that from \cite{menon} require the same number of iterations to reach convergence, while the greedy interference avoidance and the greedy MSE minimization reach the maximum allowed number of iterations.

\begin{figure}[t]
\centering
\includegraphics[width=13cm]{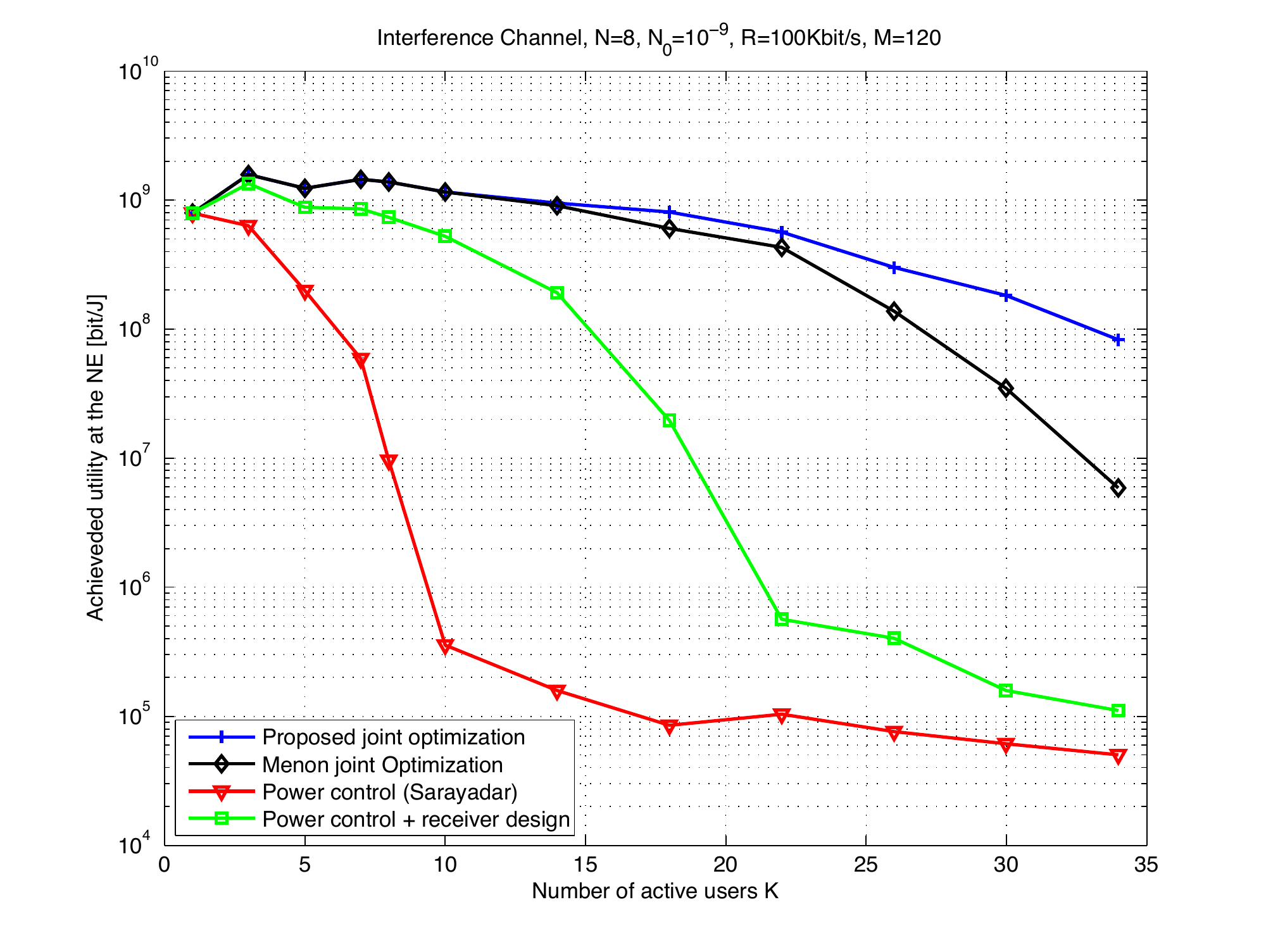}
\caption{Peer to Peer channel. $N=8$. Achieved energy efficiency (bit/Joule) at the NE versus the number of active users for four different non-cooperative games, i.e. (a) Algorithm \ref{Alg:EnergyEfficiency}, (b) Modified version of Algorithm \ref{Alg:EnergyEfficiency} that employs \cite{menon} for spreading code allocation, (c) joint power control and uplink receiver design, (d) power control with a matched filter at the receiver \cite{mandamulti}.}\label{Fig3}
\end{figure}

\begin{figure}[t]
\centering
\includegraphics[width=13cm]{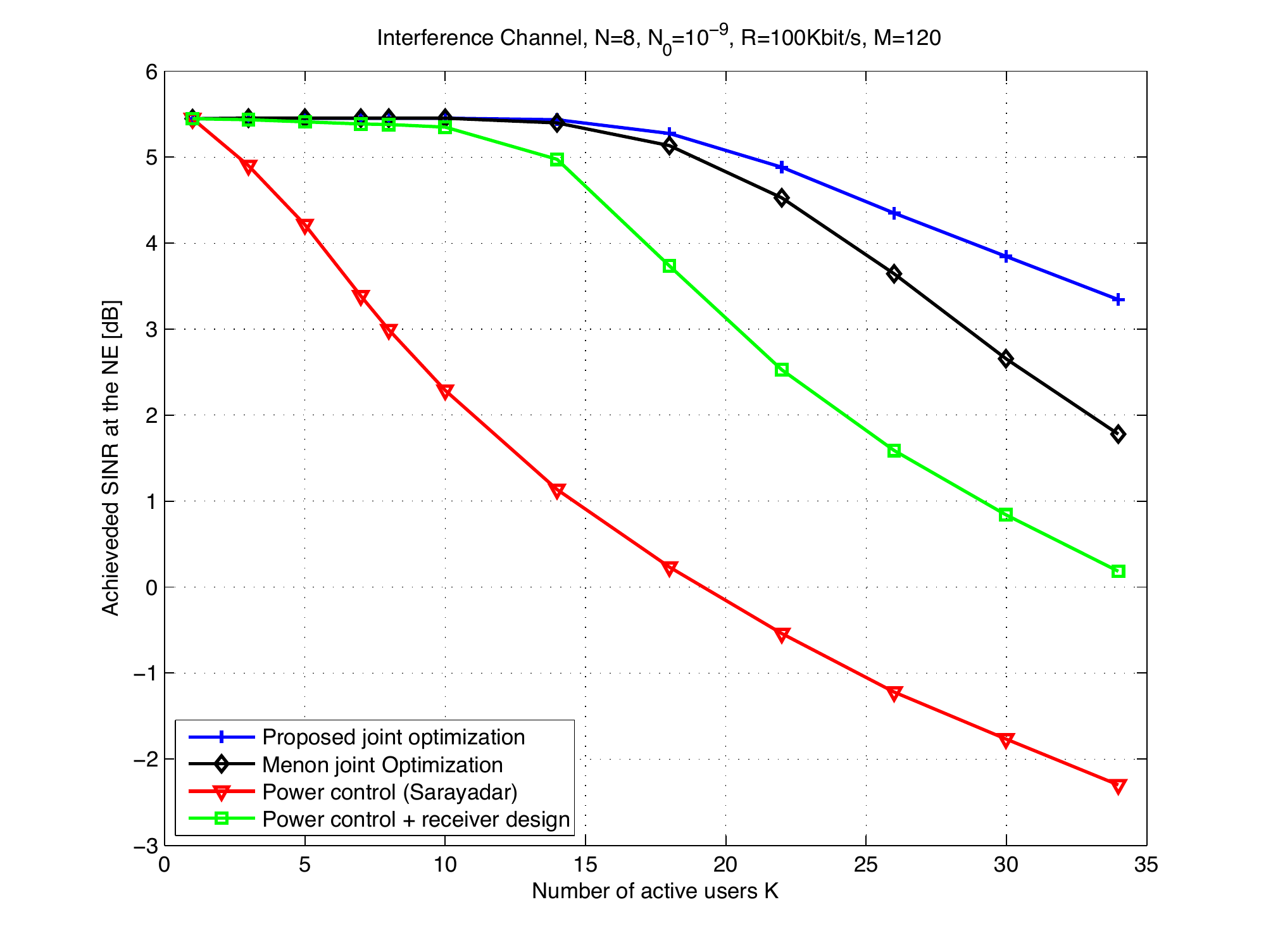}
\caption{Peer to Peer channel. $N=8$. Achieved SINR at the NE versus the number of active users for four different non-cooperative games, i.e. (a) Algorithm \ref{Alg:EnergyEfficiency}, (b) Modified version of Algorithm \ref{Alg:EnergyEfficiency} that employs \cite{menon} for spreading code allocation, (c) joint power control and uplink receiver design, (d) power control with a matched filter at the receiver \cite{mandamulti}.}\label{Fig4}
\end{figure}

In Figs. \ref{Fig3} and \ref{Fig4} we address the performance of Algorithm \ref{Alg:EnergyEfficiency} by contrasting the following algorithms.
\begin{enumerate}
\item
The proposed Algorithm \ref{Alg:EnergyEfficiency}.
\item
For the sake of comparison, we included the performance of a modified version of Algorithm \ref{Alg:EnergyEfficiency} where the spreading code allocation in STEP 1 is not implemented according to the newly proposed procedure of Section \ref{Sec:PotentialGameTMSE}, but is instead implemented according to the spreading allocation game from \cite{menon} by Menon et alii, that we outlined in Section \ref{Sec:Menon}. For this reason, this algorithm has been labeled as ``Menon joint optimization'' in the presented illustrations.
\item
In order to address the gain granted by spreading code optimization, we consider the performance obtained with power control and uplink receiver design, but no spreading code optimization.
\item
Power control with a matched filter at the receiver \cite{mandamulti}.
\end{enumerate}
Fig. \ref{Fig3} shows the achieved utility (\ref{Eq:EnergyEfficiency}) at the NE versus the number of active links. The results indicate that the proposed algorithm grants a huge performance gain with respect to other algorithms for energy efficiency. In particular the gain increases with the number of users, implying that the proposed technique is more resilient to heavy multi-user interference scenarios than its competitors.
In Fig. \ref{Fig4} a similar scenario is considered, with the difference that the achieved SINR at the NE is shown. Again, we see that the proposed method achieves better performance than other solutions.

In Tab. \ref{tab:1}, the computational complexity of Algorithm 1 is addressed in terms of number of iterations needed to reach convergence. In particular, defining the vector $\bp(n)=[p_{1}(n),\ldots,p_{K}(n)]$ containing the users' transmit powers at the $n$-th iterations of Algorithm 1, Tab. 1 shows the quantity $E(n)=\frac{\|\bp(n)-\bp(n-1)\|}{\|\bp(n)\|}$ versus $n$, for a number of users $K=3;10;25;30$.
Note that $E(n)$ is the normalized error between the power vector at the $n$-th iteration with respect to the previous iteration. Convergence in our simulations was declared when $E(n)<10^{-3}$. The results indicate that a very low number of iterations is needed to reach convergence even in overloaded scenarios in which $K$ is a little bit higher than the processing gain. Of course, the required number of iterations increases, but is still satisfactory, in heavily loaded scenarios in which the number of users is much higher than the processing gain.

\subsection{Uplink of a multi-cell wireless network}
We have considered a multi-cell wireless network with $B=4$ APs. It is assumed that each user's data are decoded at the AP with the largest channel coefficient, namely $a(k)=\arg \ds \max_{\ell=1, \ldots, B} \left(h^2_{k, \ell}\right)$.

\begin{figure}[t]
\centering
\includegraphics[width=13cm]{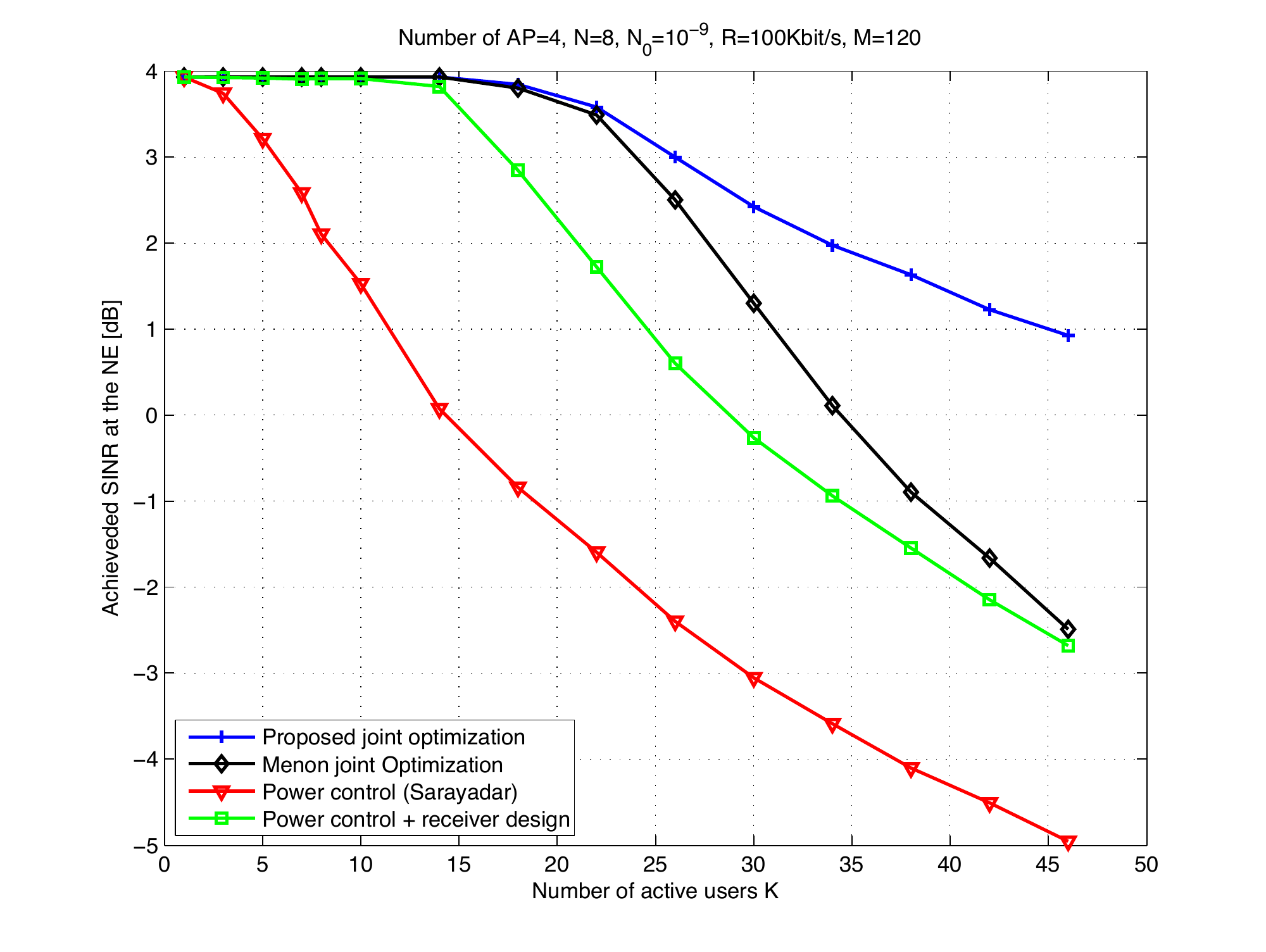}
\caption{Uplink of a multicell network. $N=8$. $B=4$. Achieved SINR at the NE versus the number of active users for four different non-cooperative games, i.e. (a) Algorithm \ref{Alg:EnergyEfficiency}, (b) Modified version of Algorithm \ref{Alg:EnergyEfficiency} that employs \cite{menon} for spreading code allocation, (c) joint power control and uplink receiver design, (d) power control with a matched filter at the receiver \cite{mandamulti}.}\label{Fig5}
\end{figure}

\begin{figure}[t]
\centering
\includegraphics[width=13cm]{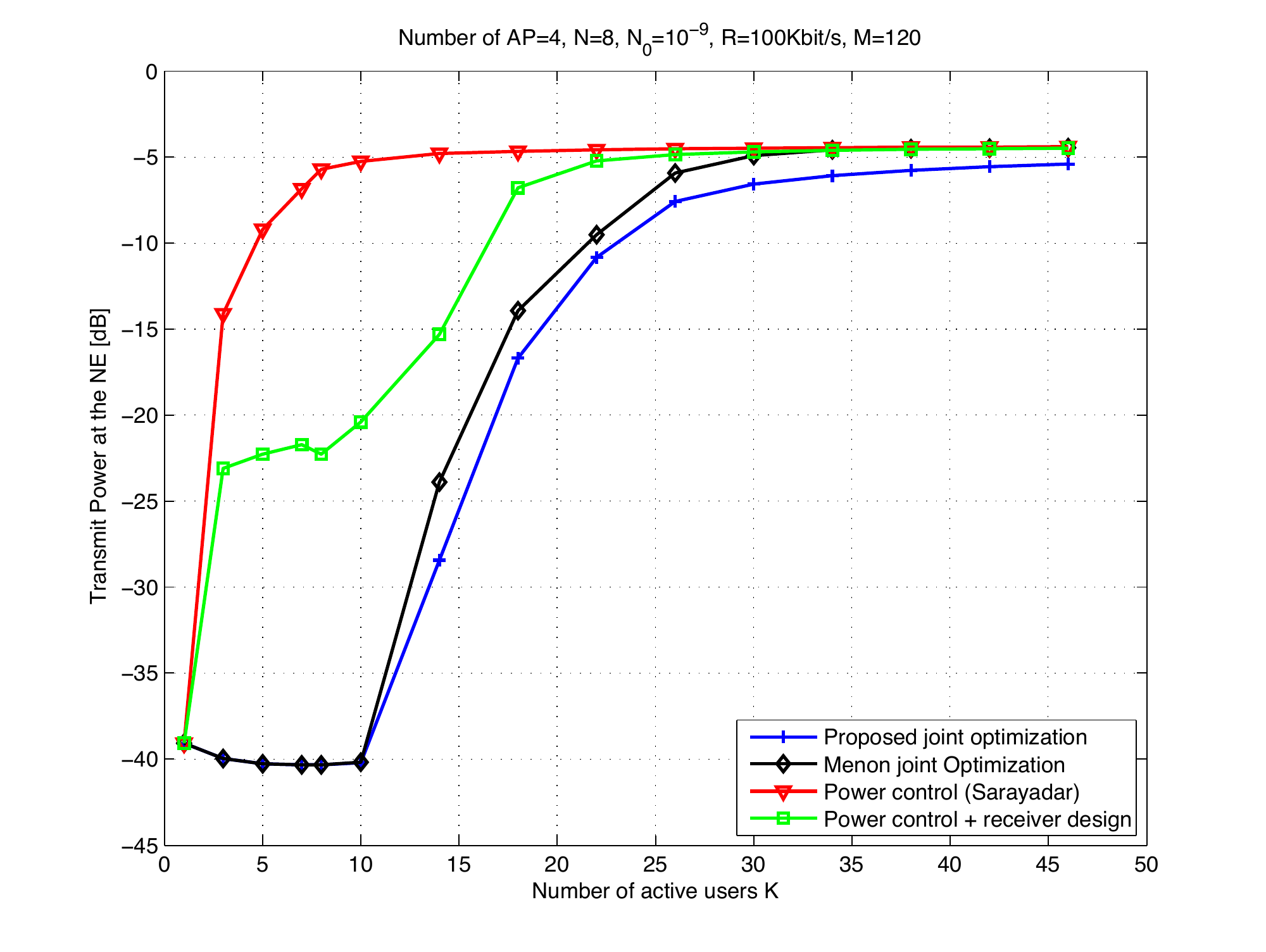}
\caption{Uplink of a multicell network. $N=8$. $B=4$. Average transmit power at the NE versus the number of active users for four different non-cooperative games, i.e. (a) Algorithm \ref{Alg:EnergyEfficiency}, (b) Modified version of Algorithm \ref{Alg:EnergyEfficiency} that employs \cite{menon} for spreading code allocation, (c) joint power control and uplink receiver design, (d) power control with a matched filter at the receiver \cite{mandamulti}.}\label{Fig6}
\end{figure}

In Fig. \ref{Fig5} and \ref{Fig6} the efficiency of Algorithm \ref{Alg:EnergyEfficiency} is addressed by contrasting the same algorithms as for the peer to peer channel case. In Fig. \ref{Fig5} the achieved SINR at the NE is shown, whereas in Fig. \ref{Fig6} the average transmit power at the NE is plotted. From the inspection of these two figures, it can be concluded that the newly proposed algorithm for energy-efficiency achieves a higher SINR than its competitors, while requiring a lower transmit power.

As for the number of iterations needed for Algorithm \ref{Alg:EnergyEfficiency} to reach convergence, similar results as in Tab \ref{tab:1} have been obtained, but we omit details for the sake of brevity.

\subsection{Uplink of a multi-cell wireless network with femtocells}
We now focus on the comparison between the multicell scenario in which there are 2 APs, and the scenario in which we have 2 APs and 4 femtocell APs serving an area of radius 100m.
The channel coefficients $h^2_{i,j}$ have been generated according to an exponential distribution with mean equal to $d_{i,j}^{-2}$, with $d_{i,j}$ the distance between the $i$-th user and the $j$-th access point. Again, it is assumed that each user's data are decoded at the AP with the largest channel coefficient, namely $a(k)=\arg \ds \max_{\ell=1, \ldots, B} \left(h^2_{k, \ell}\right)$.

\begin{figure}[t]
\centering
\includegraphics[width=13cm]{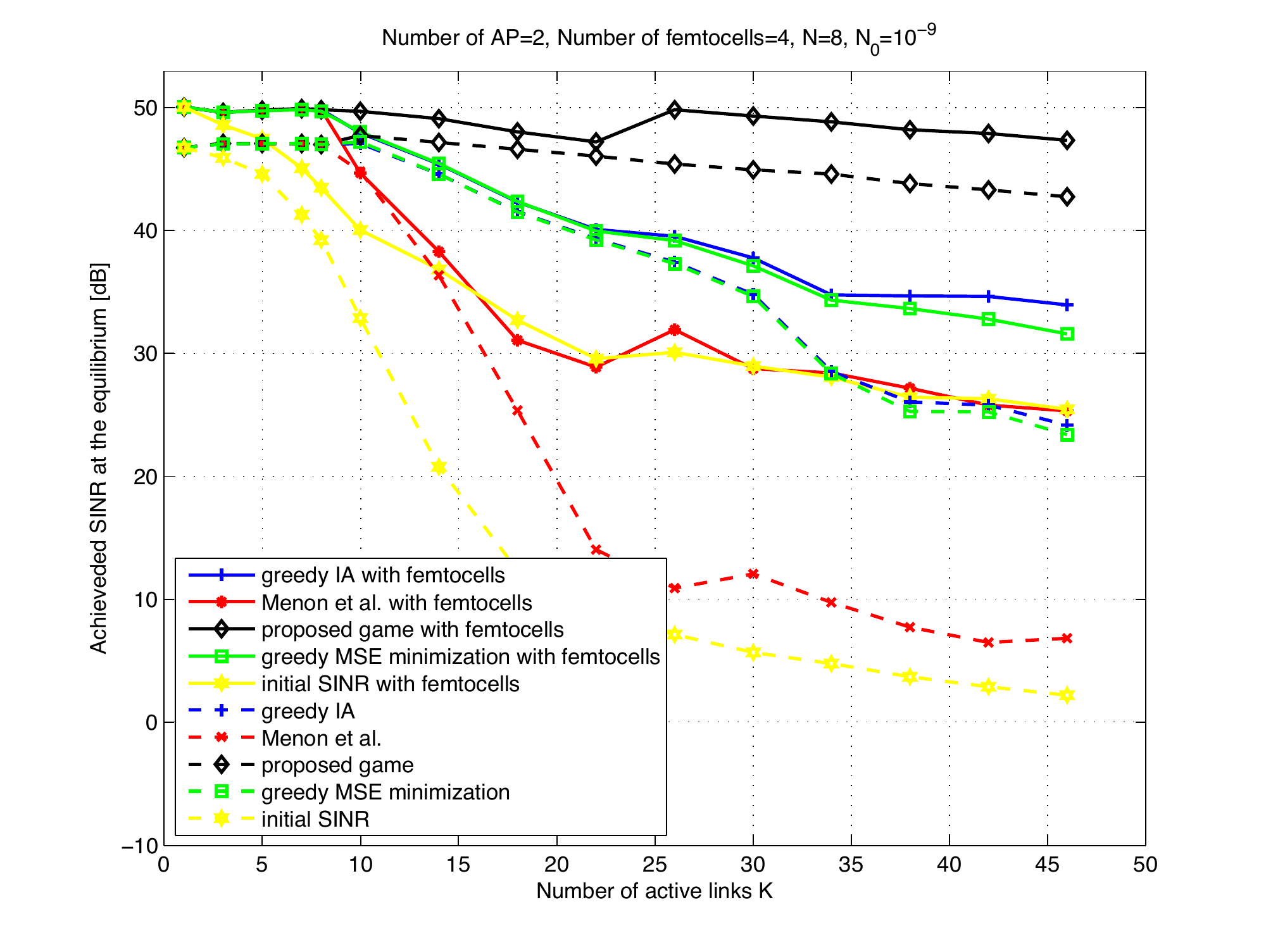}
\caption{Uplink of a multicell network with femtocells. $N=8$. $B=2$. $4$ Femtocell APs. Achieved SINR at the NE for the considered spreading code allocation procedures versus the number of active users.}\label{Fig7}
\end{figure}

First of all, we consider the waveform adaptation games discussed in Section \ref{Sec:SpreadingCodeDesign}. Fig. \ref{Fig7} shows the achieved SINR at the NE for the same algorithms considered in the peer to peer channel case, versus the number of active users. Also in this case, a maximum of 5000 iterations has been included in the simulation program in order to have a stopping rule for the resource allocation games of Sections \ref{Sec:GreedyIA} and \ref{Sec:greedyMSE}, and again it is seen that all the algorithms exhibit similar performance when $K\leq N$, whereas the newly proposed resource allocation strategy of Section \ref{Sec_IArevisited} achieves the best performance when $K>N$, and similar remarks as those made for the peer to peer channel hold. It is also seen that when femtocells are active, much better performance are obtained, and hence much more transmit energy can be saved.

\begin{figure}[t]
\centering
\includegraphics[width=13cm]{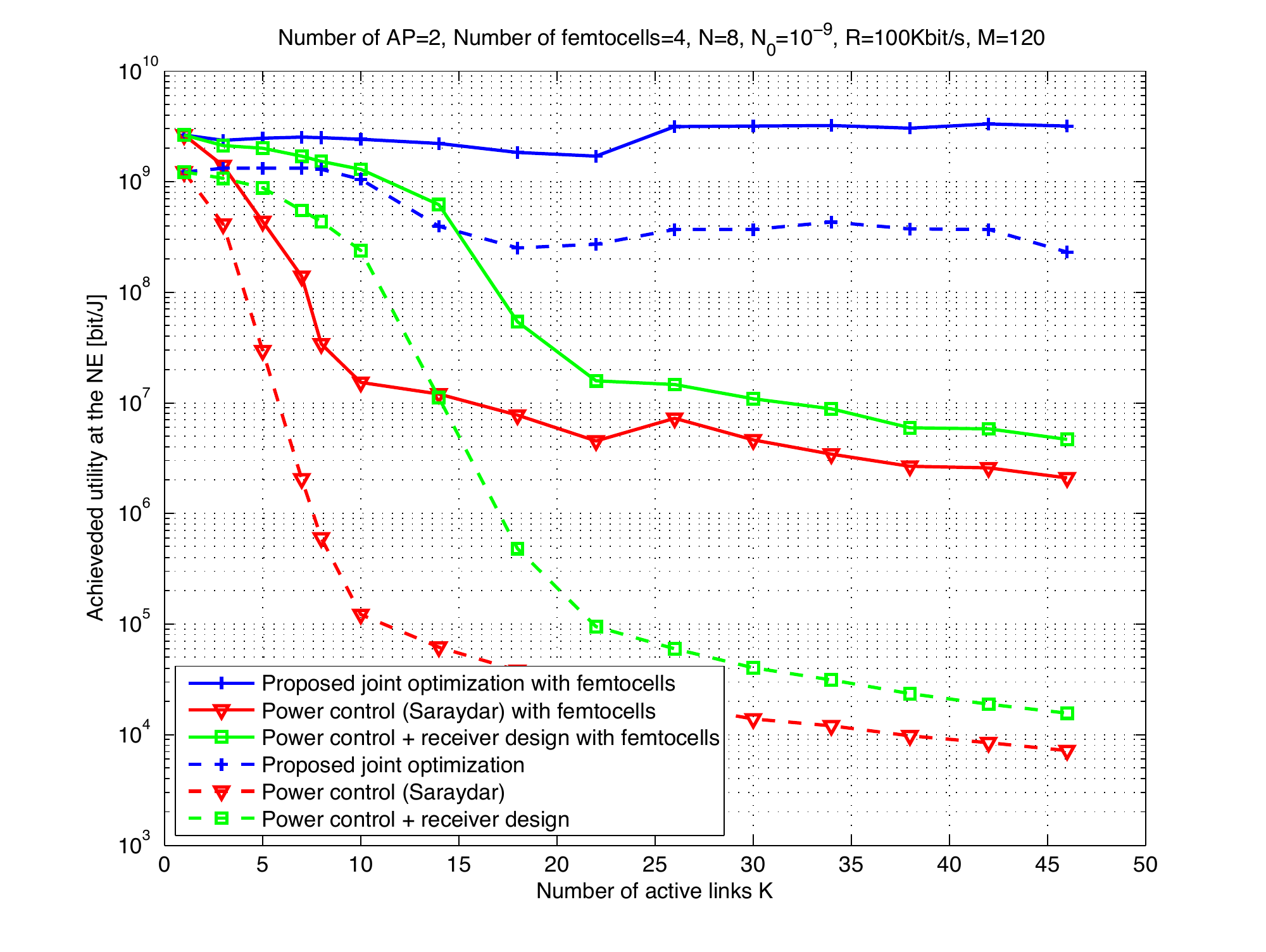}
\caption{Uplink of a multicell network with femtocells. $N=8$. $B=2$. $4$ Femtocell APs. Achieved energy efficiency (bit/Joule) at the NE versus the number of active users for three different non-cooperative games, i.e. (a) Algorithm \ref{Alg:EnergyEfficiency}, (b) joint power control and uplink receiver design, (c) power control with a matched filter at the receiver \cite{mandamulti}.}\label{Fig8}
\end{figure}

Next, in Fig. \ref{Fig8} we compare the performance of Algorithm \ref{Alg:EnergyEfficiency} to the case in which spreading code optimization is not carried out, and to the case in which only power control is performed, \cite{mandamulti}. Fig. \ref{Fig8} shows the achieved utility at the NE versus the number of active users. Again, we see that the newly proposed joint procedure greatly outperforms the competing alternatives, and that femtocells bring substantial (up to ten-fold) performance improvements.

\section{Conclusions}
This paper has considered the problem of energy-efficient non-cooperative resource allocation  in a multipoint-to-multipoint multiuser wireless data network. Leveraging on the study \cite{menon}, wherein it has been revealed that the theory of potential games can be used to obtain convergent non-cooperative resource allocation games in multi-cell networks, we have proposed resource allocation procedures that have been shown to outperform
competing alternatives. Although our paper deals with the CDMA access technique, the same concepts and the potential games framework may be applied, with ordinary efforts, to different multiple access strategies, such as orthogonal frequency division multiple access (OFDMA) strategy,
to obtain noncooperative transmit power control and carrier allocation procedures. These topics  form the object of ongoing research.

\section*{Acknowledgements}
This work was supported by the European Commission
in the framework of the FP7 Network of Excellence
in energy-efficient network design TREND, under grant number 257740.





\bibliographystyle{model1a-num-names}
\bibliography{<your-bib-database>}







\begin{table}[htbp]\centering\caption{$E(n)$ achieved by Algorithm \ref{Alg:EnergyEfficiency}, for $K=3;10;25;30$. System parameters: $N=8$. Convergence was declared when $E(n)<10^{-3}$} \label{tab:1}
\begin{tabular}{c| c | c | c | c |}
$K$ & $3$ & $10$ & $25$ & $30$\\
\hline
\hline
$n=1$ & $783$ & 284 & 4.37 & 2.31\\
$n=2$ & 0.7688 & 0.6682 & 0.0233 & 0.0886\\
$n=3$ & 0.0180 & 0.0076 & 0.0088 & 0.0038\\
$n=4$ & $2.06\times 10^{-6}$ & 0.0046 & 0.0044 & 0.0150\\
$n=5$ & -- & $1.92\times 10^{-4}$ & 0.0027 & 0.0107\\
$n=6$ & -- & -- & 0.0020 & 0.0073 \\
$n=7$ & -- & -- & 0.0016 & 0.0059\\
$n=8$ & -- & -- & 0.0011 & 0.0040\\
$n=9$ & -- & -- & $7.9\times 10^{-4}$ & 0.0036\\
$n=10$ & -- & -- & -- & 0.0023\\
$n=11$ & -- & -- & -- & 0.0012\\
$n=12$ & -- & -- & -- & $8.26\times 10^{-4}$\\
\hline
\end{tabular}
\end{table}

\end{document}